\documentclass[12pt]{iopart}
\usepackage{graphicx}
\usepackage{float}

\begin{document}

\title[Hirsch index as a network centrality measure]{Hirsch index as a network centrality measure}

\author{Monica G. Campiteli}
\address{Faculdade de Farm\'acia de
Ribeir\~ao Preto, Universidade de S\~ao Paulo, Avenida dos Bandeirantes 3900, 14040-901,
Ribeir\~ao Preto, SP, Brazil}
\author{Adriano J. Holanda, Paulo R. C. Soles, Leonardo H. D. Soares, Osame Kinouchi}
\address{Departamento de F\'isica e Matem\'atica, FFCLRP, Universidade de S\~ao Paulo, 
Avenida dos Bandeirantes 3900, 14040-901, Ribeir\~ao Preto, SP, Brazil}

\begin{abstract}

We study the $h$ Hirsch index as a local node centrality measure 
for complex networks in general. The $h$ index is compared with the Degree
centrality (a local measure), the Betweenness and Eigenvector centralities (two non-local
measures) in the case of a biological network (Yeast interaction protein-protein network)
and a linguistic network (Moby Thesaurus II). In both networks, the
Hirsch index has poor correlation with Betweenness centrality 
but correlates well with Eigenvector centrality, 
specially for the more important nodes that are relevant for ranking purposes, say in
Search Engine Optimization. In the thesaurus network, the $h$ index seems even to outperform
the Eigenvector centrality measure as evaluated by simple linguistic criteria. \end{abstract}


\maketitle

The Hirsch or $h$ index has been proposed and mainly studied as a scientific productivity 
statistics, being applied to individual
researchers~\cite{Hirsch:2005,Hirsch:2007,Batista:2006,Bornmann:2007,Alonso:2009}, 
groups~\cite{vanRaan:2006}, journals~\cite{Braun:2006,Bollen:2009} 
and countries~\cite{Csajbok:2007} using data from citation networks. 
In this context, a researcher has Hirsch index $h$ if he/she has 
$h$ papers with at least $h$ citations each~\cite{Hirsch:2005}. 

Recently, Korn {\em et. al}~\cite{Korn:2009} have proposed a general
index to network node centrality based on the $h$-index. 
{\em Korn et al.} named it as the {\em lobby index}, but since it is simply the application of 
Hirsch's idea in the context of general networks, we shall continue to call it the Hirsch 
(centrality) index.  {\em Korn et al.} argue that 
the proposed index contains a mix of properties of other well known centrality measures. 
However, they have studied it mostly  in the context of artificial or idealized networks 
like the Barabasi-Albert model \cite{Korn:2009, Barabasi:1999b}. 

Here we study the Hirsch index in two real life networks 
and discuss some computational 
and conceptual advantages of the $h$-index as a new centrality measure. 
We study the Hirsch centrality in
linguistic and biological networks already considered by the physics community. 
The first one is the {\em Moby Thesaurus II} network~\cite{MobyII,Motter:2002,Holanda:2004} 
composed by $30,260$ nodes and around $1.7$ million links. 
The biological network is the yeast protein-protein network obtained from the
\textit{Biogrid} repository \cite{Biogrid}. 
This is a curated repository for physical and genetic interactions for $5,433$ proteins and over
$150,000$ unambiguous interactions.
The advantage of using these networks as benchmarks to evaluate node centrality is that they enable 
us not only a comparison with other centrality indexes but also an independent evaluation by 
standard linguistic/biological  criteria.

In this work we use the following definition:
\begin{quote}
{\em The Hirsch centrality index $h$ of a node is the largest integer 
$h$ such that the node has at least $h$ neighbours which have a degree of at least $h$.}
\end{quote}

The linguistic network is 
formed by the entries or "root words" of the {\em Moby Thesaurus II}~\cite{MobyII}. 
To construct the network we use the convention that an outlink goes from a 
root word to a related word. The raw thesaurus have
over $2.5$ million links but there are many words with only in-links (that is, they are not
root words). So, we worked with a cleansed version with around $1.7$ million links 
where only root words constitute nodes~\cite{Motter:2002,Holanda:2004}.
The minimal number of outlinks is $17$ and the maximum is $1106$. Notice that
the graph is directed~\cite{Holanda:2004} but we have used as centrality measures the out-degree 
and the $h$-index based on the out-degree (from here referred simply as ''degree" $D$).

The biological network as downloaded from the \textit{Biogrid} is composed by gene 
products connected by a link\cite{Biogrid}. The links include direct physical binding of two proteins, 
co-existence in a stable complex or genetic interaction as given by one or several 
experiments described in the literature. The links are considered here undirected. 
The degree distribution ranges from $1$ to $1,975$.

\begin{figure}[H]
 \centering
\includegraphics[width=0.8\columnwidth]{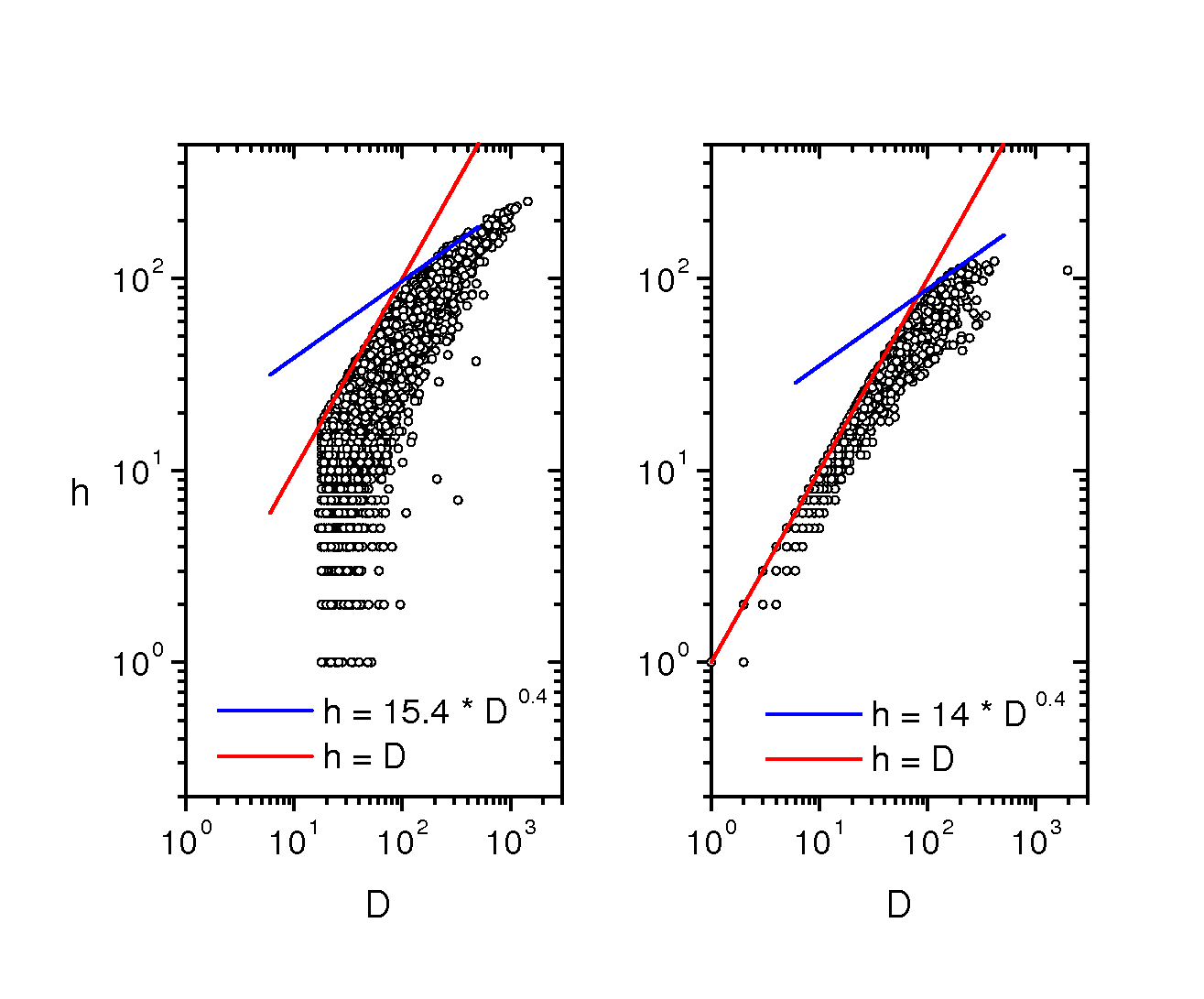}
 \caption{Log-log dispersion plot of $h$ versus Degree centrality $D$ for a) Moby Thesaurus II 
 and b) Yeast network.
   \label{F1}}
 \end{figure}

In Figure~\ref{F1} we present dispersion plots of the $h$-index versus degree $D$ for the two networks.
We notice that, although correlated, the two measures are not redundant. 
More importantly, in the thesaurus case the
$h$-index highlights false positives (defined in terms of their degree centrality), 
that is, words with high degree but low $h$.
We also note that, by definition, a node cannot have $h>D$, and that the boundary $h = D$ is 
saturated in the low $D$ regime (up $D=100$).
For higher $D$, we observe
in both networks that the highest $h$ are proportional to $D^{0.4}$, but the origin of 
this anomalous exponent is not clear.

\begin{figure}[H]
\centering
 \includegraphics[width=0.8\columnwidth]{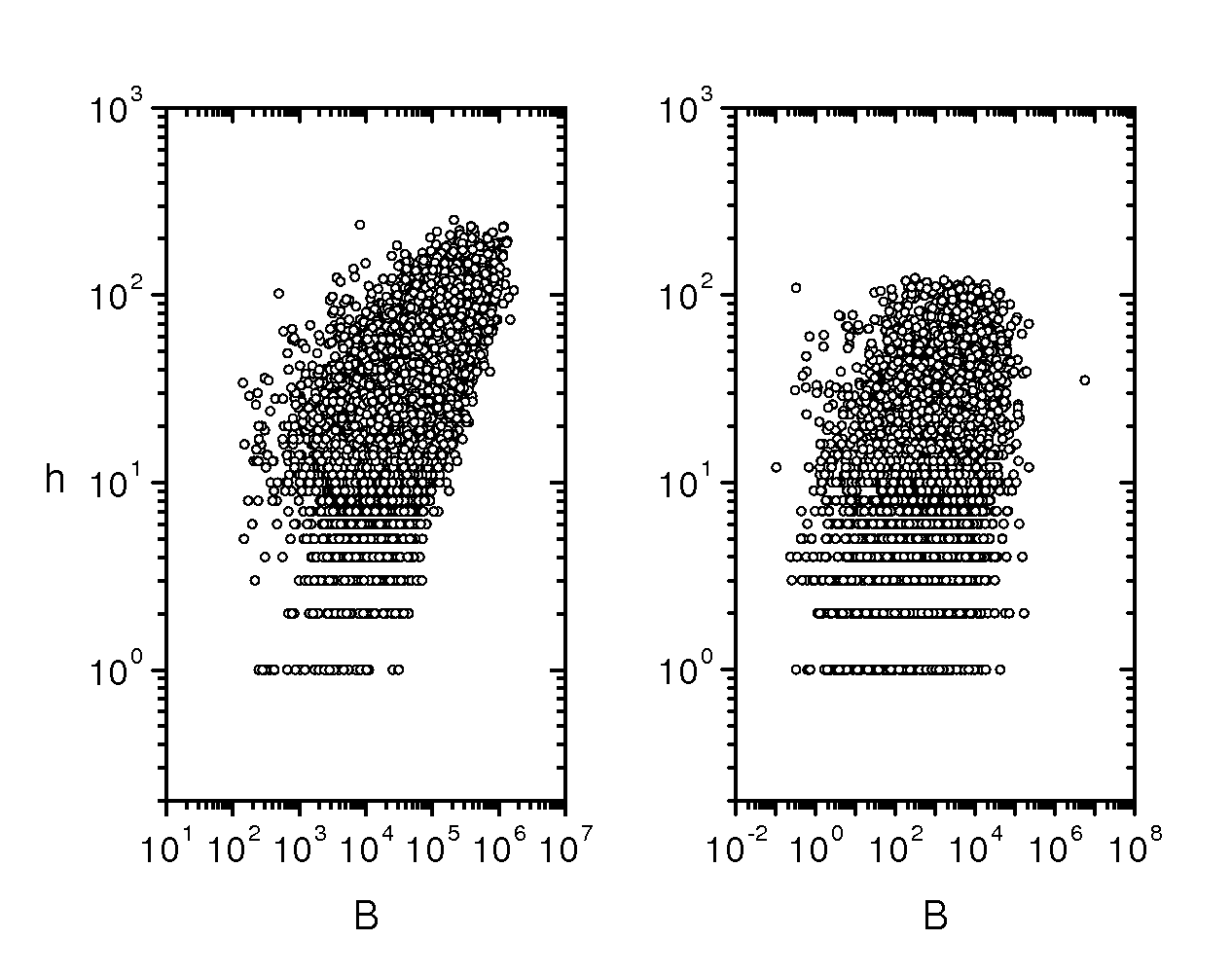} 
 \caption{Log-log dispersion plot of $h$ versus Betweenness $B$ for 
a) Moby Thesaurus II network and b) Yeast network.
   \label{F2}}
 \end{figure}

Now we compare the Hirsch index with two standard non-local centrality indexes, 
Betweenness and Eigenvector centrality. First, we present
in Figure~\ref{F2} the dispersion plots of $h$ versus Betweenness centrality $B$. 
No strong correlation is apparent, meaning that the indexes seems to contemplate different ideas of
centrality (this will be discussed better bellow).

 \begin{figure}[H]
 \centering
 \includegraphics[width=0.8\columnwidth]{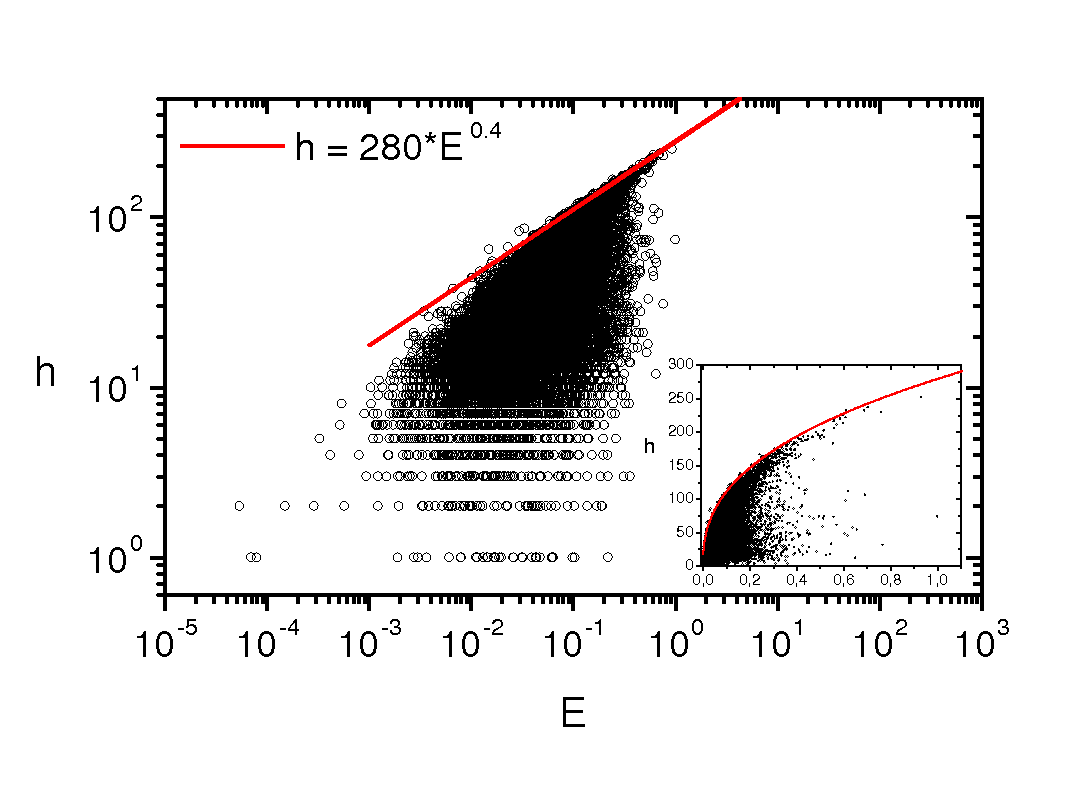}
 \caption{Log-log dispersion plot $h$ versus Eigenvector centrality $E$ 
for the Moby Thesaurus II. Inset: Linear scale, 
notice the several words with high E but low h. \label{F3}}
 \end{figure}

In Figure~\ref{F3} we give the dispersion plot for the $h$-index versus the Eigenvector 
centrality $E$ for the thesaurus network. 
In the high $E$ regime the maximal $h$ values seem to be bounded by $h \propto E^{0.4}$, 
like in the $h$ versus $D$ plot.
We observe several nodes with high $E$ but relatively low $h$ (see Inset).
Examining individually these nodes, we find that $h$ seems to outperform $E$ 
in the ranking task, since words with
high $h$ also have high $E$ and are basic and important polysemous words. 
In contrast, terms
with high $E$ can have high or low $h$. Those with low $h$ are mostly phrasal 
verbs or multiple word expressions derived from the words with high $h$.
 
It is difficult to quantify the quality of some ranking list, but the above effect is very clear,
as can be observed in Table~\ref{Table} that shows the top $25$ words 
ranked by $h$ and $E$ (the same occur for other high $E$ and low $h$ words).

\begin{table}[ht!]

\caption{\label{Table} Top 25 words ranked by Hirsch centrality (left) and by
Eigenvector centrality (right).}	\begin{center}
	\begin{tabular}{c|c|c|c|c|c|c}
	 \textbf{$h$ rank} & && & \textbf{$E$ rank} & & \\ \hline
  \textbf{Hirsch} & \textbf{Eigenvector} & \textbf{Word} & $\:\:\:$& \textbf{Eigenvector} & \textbf{Hirsch} 
  & \textbf{Word} \\ \hline
  252 & 0,930	&  cut    & & 1,000 & 74 & cut up\\
	237 & 0,701	&  set    & & 0,930 & 252 & cut\\
	233 & 0,608	&  run    & & 0,765 & 31 & set upon\\
	232 & 0,687	&  line   & & 0,760 & 230 &turn\\
	230 & 0,760	&  turn   & & 0,701 & 237 & set\\
	225 & 0,598	&  point  & & 0,690 & 106 & break up\\
	222 & 0,608	&  cast   & & 0,687 & 232 & line\\
	220 & 0,584	&  break  & & 0,656 & 54 & line up\\
	218 & 0,560	&  mark   & & 0,649 & 12 & run wild\\
	216 & 0,558	&  measure& & 0,637 & 57 & turn upside down\\
	213	& 0,597 &  pass   & &   0,618 & 112& make up \\
  211 & 0,570 &  check  & &   0,617 & 45 & cast up \\
  209 &	0,487 &  crack  & &   0,608 & 222& cast \\
  206 & 0,562 &  make   & &   0,608 & 233& run \\
  203	& 0,448 &  dash  & &   0,608 &  97& crack up \\
  203 & 0,517 &  stamp   & &   0,604 &  48& check out \\
	202 & 0,514 &  work   & &   0,598 & 225& point \\
	200 & 0,484 &  strain & &   0,597 & 213& pass \\
	196 & 0,491 &  hold   & &   0,584 & 220& break \\
	195 & 0,508 &  form   & &   0,571 &  61& pass up \\
	194 & 0,447 &  beat   & &   0,570 & 211& check \\
	193 & 0,500 &  get    & &   0,562 & 206& make \\
	193 & 0,429 &  rank  & &   0,560 & 218& mark \\
	193 & 0,469 &  round   & &   0,558 &  73& fix up \\
	192 & 0,517 &  go     & &   0,558 & 216& measure \\
	
	\end{tabular}	
	\end{center}
\end{table}

\begin{figure}[H]
\centering
 \includegraphics[width = 0.8\columnwidth]{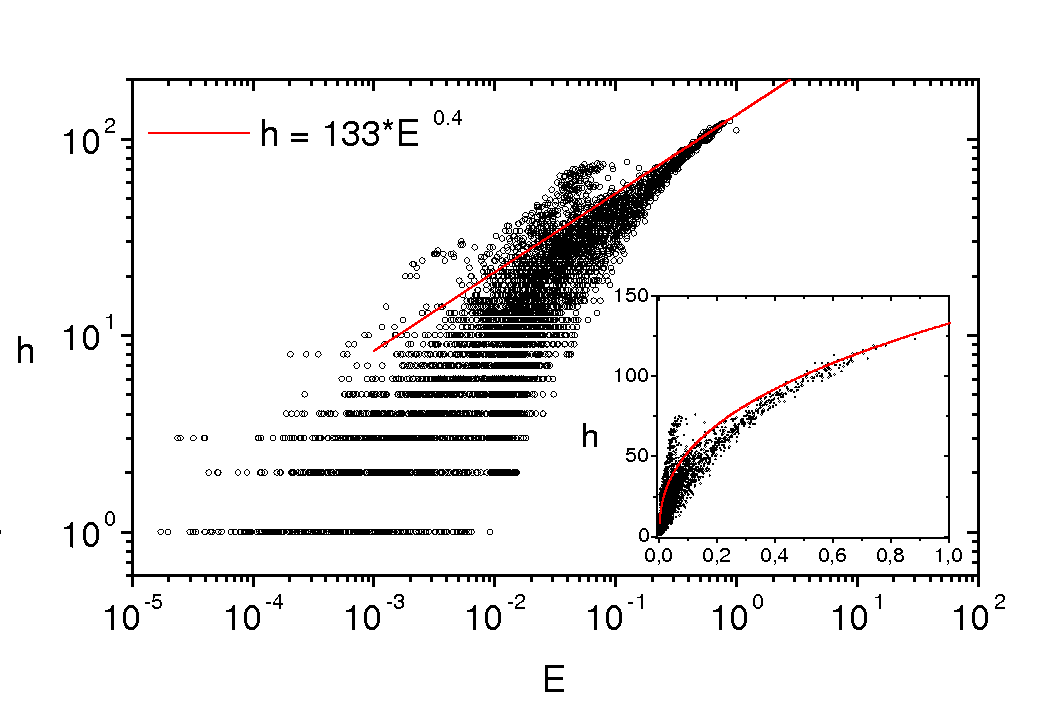}
\caption{\label{F4}Log-Log dispersion plot of $h$ versus $E$ for the Yeast network. 
 The $h$ and $E$ centralities are well correlated for $E > 0.2$ where there is a 
 $h \propto E^{0.4}$ bound for the highest $h$ values. Inset:
linear scale, notice the cluster of high $h$ but low $E$ ribosome proteins.} 
\end{figure}

In the case of the Yeast protein network (see Figure~\ref{F4}) we observe a strong correlation 
between $h$ and $E$ for $E > 0.2$. 
We notice that this regime is the relevant one for ranking purposes, say in the ranking of 
WWW pages or, in our case, to detect the most important proteins.
The highest $h$ seem also to be bounded by a $h \propto E^{0.4}$ behavior.

We also observe a detaching cluster of nodes with low $E$ and
moderate $h$ (see Figure~\ref{F4}). It is very interesting that all these nodes seem to pertain to
ribosome proteins, meaning that the $h$ index 
carries information that can be useful for detecting modules of functionally related proteins. 
This will be studied in detail elsewhere.

So, the important point is that, in the regime relevant for ranking purposes, 
the biological network data shows a 
strong correlation between the Eigenvector and Hirsch centralities, although
the computation of the Hirsch index is much less demanding because it is not iterative and
uses only local information. This suggests that the
$h$ centrality can be useful for ranking purposes in large databases 
with results comparable with Eingenvector centrality. 
This claim could be tested in the paper citation network studied by 
Chen {\em et al.}~\cite{Chen:2007} where the \textit{PageRank} algorithm, which core is the
Eigenvector centrality concept, has given interesting results.

Now we argue that a local centrality measure like the $h$-index makes more sense for
some kinds of networks than global measures like Betweenness and Closeness~\cite{Newman:2001}. 
The Betweenness of a given node is proportional to the number of "geodesical" paths 
(minimal paths between node pairs in the network) that pass through it.
Similarly, the Closeness  of a node is the average length of the minimal paths between such
node and all other nodes of the network. 
These seems to be important measures for networks where such minimal paths represent
transport channels for information (internet, social networks), 
energy (powergrids), materials (airports network) or diseases (social and sexual networks).

Both centrality measures make sense for studying 
diffusion and epidemic processes in transport networks, but
the relevance of minimal paths is not so clear for linguistic or cultural 
networks like thesauri or, as another example, the network of
culinary ingredients studied in~\cite{Kinouchi:2008} where links represent associations but not channels.
For networks similar to the linguistic one studied here there is a strong decay
of correlations: two words  $A$ and $C$ with minimal path of two 
links (that is, $A-B-C$) are almost uncorrelated, since this means that $C$ is not a 
word semantically related to $A$. 
The paths between words may be relevant to describe perhaps associative psychological processes 
(say, $A$ remembers $B$ that remembers $C$), but they are not channels in the same sense 
of physical transport networks. So, the locality of the $h$-index could be an advantage
to its application for ranking nodes in non-transport networks 
where path distance or channel flux 
has poor relevance and are not important aspects to define centrality~\cite{Borgatti:2005}. 
We notice that this could be the case of WWW pages since links represent more associations than
channels and users do not navigate from link to link by large distances. 

In conclusion, we studied the $h$-index~\cite{Korn:2009} in the Moby II 
Thesaurus network and the protein-protein interaction Yeast network. 
Several characteristic of this new centrality index have been highlighted. 
The Hirsch index seems to be a better local 
index than the node degree $D$ because it incorporates information about the importance of 
the node neighbours. Being local, it is computationally cheap to calculate.
It seems to make more sense for non-transport networks than
some global measures.  

We also found that the $h$-index is more correlated to
Eigenvector centrality than Betweenness centrality. Indeed, in the ranking task for words
in the thesaurus, Hirsch centrality seems even to outperform the 
Eigenvector measure as a centrality index,
detecting basic polysemous words instead of more complicated and rare terms.
Since Eigenvector centrality corresponds to the
core idea behind the original \textit{PageRank} algorithm~\cite{Chen:2007}, 
which is very computationally demanding at the Internet scale, we suggest that the $h$-index could 
furnish auxiliary information for ranking pages in the area of Search Engine Optimization.
Due to the simplicity of the definition and computation of the $h$-index, comparing its 
performance with standard global centrality measures in other physical, biological and social networks 
promises very interesting results. 

\ack{
This work received financial support from CNPq and FAPESP. 
The authors are grateful to N. Caticha, R. Vicente, A. S. Martinez, R. Rotta, P. D. Batista and T. Penna 
for discussions and suggestions.}

\section*{References}

\bibliography{myrefs}
\bibliographystyle{unsrt}

\end{document}